\documentclass[aps,prB,amssymb,longbibliography,superscriptaddress,twocolumn]{revtex4-1}
\usepackage{graphicx}
\usepackage{bm,color}

\usepackage{amsmath}
\usepackage{bm,color}
\usepackage{multirow}
\usepackage{ulem}
\usepackage{xcolor}

\begin{document}

\title{Topological pump and bulk-edge-correspondence in an extended Bose-Hubbard model}
\date{\today}

\author{Yoshihito Kuno}
\author{Yasuhiro Hatsugai}
\affiliation{Department of Physics, University of Tsukuba, Tsukuba, Ibaraki 305-8571, Japan}

\begin{abstract}
An extended Bose-Hubbard model (EBHM) with three- and four-body constraints can be feasible in cold atoms in an optical lattice. 
A rich phase structure including various symmetry-protected topological (SPT) phases is obtained numerically with suitable parameter settings and particle filling. 
The SPT phase is characterized by the Berry phase as a local topological order parameter 
and the structure of the entanglement spectrum (ES). 
Based on the presence of various topological phases, separated by gapless phase boundaries, 
the EBHM exhibits various bosonic topological pumps, which are constructed by connecting the different SPT phases without gap closing. The bulk topological pumps exhibit the plateau transitions characterized by many-body Chern numbers.
For the system with boundary, the center of mass (CoM) under grand canonical ensemble elucidates the contributions of multiple edge states and reveals the topology of the system. 
We demonstrate that the interacting bosonic pumps obey the bulk-edge-correspondence.
\end{abstract}


\maketitle
\section{Introduction}
The gapped symmetry protected topological (SPT) phase \cite{Pollmann2010,Chen,Pollmann2012} is now a hot topic in the condensed matter physics.
The topological insulators (TI) are one of the typical examples.
The gap is protected by some symmetries, that do not vanish for small but finite perturbation as far as the symmetries are preserved. 
The SPT phase is robust. This is the topological protection. Also, SPT phase is characterized by the presence of edge-states. 
The bulk-edge correspondence clearly characterizes the appearance of the SPT phase \cite{Hatsugai1993} that is also stable for interaction as far as the symmetries are preserved. 
The SPT phases in the fermionic or spin models such as Haldane $S=1$ chain have been extensively studied. However, the concrete example of the bosonic SPT phase in interacting bosonic lattice models are rare 
although the bosonic SPT is predicted by the group cohomology \cite{Chen2012}. 
The demonstration of the SPT phases is essential. 
So far, in the extended Bose Hubbard model (EBHM) \cite{Dutta,Baier,Lahaye}, the Haldane insulator (HI) as an analogue of the $S=1$ Haldane phase of the quantum spin chain is investigated \cite{DallaTorre2006,Berg2008,Rossini2012,Ejima2014,Batrouni2013,Deng,Gremaud,Kawaki2017,Stumper,Fraxanet}. 
Note that such a HI phase can be realized in fermionic gas trapped in a ladder optical lattice system \cite{Fromholz}. 

Also, the study of charge \cite{Thouless} or spin \cite{Shindou} pump based on the topological phases get focused \cite{Berg2011,Rossini2013,Wang,Hatsugai2016,Nakagawa2017,RLi,Kuno2017,Hayward,Greschner2020,KH2020,KH2021}. Recent experimental development of photonic crystals and cold atoms have realized topological charge pumps \cite{Kraus_ex,Lohse,Nakajima,Schweizer} 
and demonstrated its stability for perturbations \cite{Nakajima2021}. 
On the theoretical side, a bosonic topological pump based on the HI phase has been confirmed from the bulk perspective \cite{Berg2011,Rossini2013} and the presence of topological charge pumps in some bosonic systems have been reported \cite{Wang,Nakagawa2017,RLi,Kuno2017,Hayward,Greschner2020}. Furthermore, the bulk-edge-correspondence of the topological pump has been discussed \cite{Hatsugai2016}.

In this work, we discuss topological phenomena in the EBHM by introducing the dimerization and the local Hilbert space constraints 
and found various SPT phases. 
Also, since the EBHM has a high degree of freedom due to the local particle number constraint, richer and more complex phases are expected compared with quantum spins or fermions. 
In particular, we have demonstrated the dimerization of the hopping gives various SPT phases in the EBHM. 
The appearance of SPT phases strongly depends on the local Hilbert space constraints and mean particle density. 
These SPT phases are analogue of the valence-bond-solid (VBS) states in the spin $S\geq 1$ chains \cite{Lauchli,Katsura2007,Hirano2008}. 
Compared to the spin $S \geq 1$ chains, the EBHM is simple in its algebraic structure. 
It implies the SPT of the EBHM is fundamental and more universal.
We generalize the VBS picture for the SPT of the EBHM. 
It {\it does not} correspond to that of the quantum spin in various aspects. 

In this work, we first investigate non-trivial topological phases of the EBHM. 
We numerically found the SPT phases in the EBHM under a suitable parameter set.
{The numerically obtained SPT phases are analyzed by considering trial wave functions that describe the generalized VBS. The numerically obtained SPT phases can be characterized by the Berry phase as local topological order parameter \cite{Hatsugai2005,Hatsugai2006,Hatsugai2007}. The Berry phase indicates that the numerically obtained SPT states is adiabatically connected with the generalized VBS, which cannot be decomposed into the smaller elements under the bond centered inversion symmetry.}

{Furthermore, we directly construct the topological pump by extending the parameter space since the global $U(1)$ symmetry that guarantees charge conservation is only a key factor for its construction.}  Based on the various SPT phases, by connecting the SPT phases with a symmetry-breaking term \cite{Berg2011,Rossini2013,KH2020}, 
we find the plateau transitions of the bulk topological pump. 

We further treat the open boundary case of the topological pump in detail. 
In the system with open boundary, depending on the particle filling and the dimerization parameter, the EBHM exhibits multiple edge states. 
Their energies are not fixed by the symmetry. 
These edge states play an essential role in the behavior of the center of mass (CoM) in the topological pump. 
In particular, we calculate the CoM for the grand canonical ensemble. 
It captures the contribution of edge states to the CoM for the presence of the multiple edge states. 
From the detailed study of the CoM, we confirm that the bosonic topological pumps obey the bulk-edge-correspondence that is analogous to the fermion/spin systems \cite{Hatsugai2016} in the EBHM. 

The rest of the paper is organized as follows. 
In Sec.~II, we describe the model and its basic properties. 
In Sec.~III, the presence of various topological phases is clarified. 
We introduce a generalized VBS states, which can describe features of the topological phase in the bulk and then confirm that numerically obtained states possess the features of the generalized VBS states. 
In Sec.~IV, the topological charge pumps in the EBHM are numerically demonstrated with or without edges.
Section V is devoted to the conclusion.

\section{Model}
Let us consider an extended Bose Hubbard model (EBHM), the Hamiltonian given by
\begin{eqnarray}
H_{EBH}&=&\sum^{L-1}_{j=0}\biggl[J_{j}b^{\dagger}_jb_{j+1}+{\mbox {\rm h.c.}}\nonumber\\
&&+\frac{U}{2}(n_{j}-{\bar n})^{2}+V_j (n_j-{\bar n})  (n_{j+1}-{\bar n})\biggl],
\label{EBHM}
\end{eqnarray}
where $b^{(\dagger)}_j$ is a boson annihilation (creation) operator, $n_j$ is a boson number operator $n_j=b^{\dagger}_j b_j$, the hopping dimerization is introduced by $J_j=J_1$ for $j\in$ {\rm even}, $J_j=J_2$ for $j\in$ {\rm odd}, $\bar{n}$ is a mean density per site, 
$U$ are an on-site interaction, $V_j$ is a site dependent nearest-neighbor (NN) interactions, and $L$ is the system size, which is set to be even number. 
The Hamiltonian $H_{EBH}$ is experimentally feasible in the cold atom optical lattice system \cite{Dutta}. 
The hopping dimerization $J_j$ is created by introducing an optical superlattice. The on-site $U$ and NN interactions can be independently controllable:  
the onsite $U$ is s-wave scattering, which can be controlled by using Feshbach resonance technique, 
and also $V_j$ may be tuned by employing a dipole-dipole interaction \cite{Baier,Lahaye}. 
We further introduce a generic $N$-body interaction 
\begin{eqnarray}
{\hat V}^{N}_{int}=\sum^{L-1}_{j=0}V_N\prod^{N}_{\ell=1}[n_j-(\ell-1)].
\label{V3}
\end{eqnarray}
If $V_N$ is very large ($V_N\gg U,V_j, J_j$ ($\geq 0$)), 
the local bosonic Hilbert space is trancated, 
that is, the allowed local boson number bases are restricted to $|0\rangle,\: |1\rangle, \cdots, |N-1\rangle$. 
Especially, ${\hat V}^{N=3}_{int}$ with $V_{N=3}\to \infty$, so called three-body constraint, is feasible in real experiments 
in controlling $U$ via Fechbach resonance technique \cite{Daley}. 
This three-body constraint implies $(b^{\dagger}_j)^3=0$. 
In what follows, we consider the three- and four-body constraint ($(b^{\dagger})^4=0$) separately. 

For the translational invariant case, $J_1=J_2$ and $V_j=V$, 
the global phase diagram of the Hamiltonian $H_{EBH}$ has been studied extensively \cite{DallaTorre2006,Berg2008,Rossini2012,Ejima2014,Kawaki2017}, 
where the Haldane insulator (HI) appears in the regime where $U$ and $V$ compete with mean density one. 
This HI is a SPT phase protected by the bond centered inversion symmetry, $b^{(\dagger)}_{j}\to b^{(\dagger)}_{L-1-j}$. 
It is robust against a perturbation as long as the inversion symmetry is not broken \cite{Berg2008}.

The EBHM defined in Eq.~(\ref{EBHM}) has a subtle difference from the general spin-$S$ ($N=2S+1$) chain.
Depending on the $N$-body constraint and the mean density, the EBHM is related to the dimerized spin-$S$ chain, which has a rich topological phase diagram \cite{Lauchli,Katsura2007,Hirano2008,KH2021}. 
When the local Hilbert space of boson number is trancated as $|0\rangle,\:\cdots, |n\rangle$ ($n=2S$), the boson operator is naively related to the spin-$S$ operator as $S^{-(+)}_j\sim b^{(\dagger)}_j$ and $S^{z}_j\sim n_j-{\bar n}$. 
Under this assumption, if the uniform hoppings and NN interactions are considered and the local boson occupation is truncated up to $|2\rangle$, the HI phase can be regarded as an analogous of the VBS state (Haldane phase) in the antiferromagnetic $S=1$ spin chain \cite{DallaTorre2006}. 
However, the EBHM does not exactly correspond to the spin-$S$ model such as the spin-$S$ XXZ model since the inversion symmetry in the spin space, $S^{z}_i\to 2S -S^{z}_i$ is absent for the boson counterpart.

\section{Topological phases of the bulk}
In the general spin-$S$ chain, the dimerization leads to an interesting ground state phase diagram \cite{Lauchli,Hirano2008} including various SPT phases, which are captured by VBS pictures \cite{Katsura2007,Hirano2008}.   
By the mapping between the EBHM and the generic spin-$S$ chain, we expect that the introduction of the dimerization for the hopping and NN interaction in the EBHM 
leads to rich phase diagram, especially, various SPT phases regarded as an extension of the Haldane insulator phase. 

The feature of the HI phase of the EBHM is captured by a short-range entanglement. 
This is analogous to the VBS state in the $S=1$ spin chain \cite{Berg2008} 
and higher $S$ extension as well \cite{Katsura2007,Hirano2008}.
In this section, we propose a generalised VBS as a representative states of various SPT phases in the EBHM. These states can be regarded as an “irreducible cluster state”, which cannot be decomposed into the smaller elements under the bond centered inversion symmetries which protect the topological phases.

Furthermore, we employ the Berry phase to characterize various SPT phases numerically obtained in the EBHM. The quantized Berry phase has been used for characterizing various SPT phases in many quantum many-body systems \cite{Hatsugai2005,Hatsugai2006,Hatsugai2007,Hatsugai2011,PRL-TK-TM-YH,Hirano2008,Mila,Fubasami,Araki}. The key observation is that 
Berry phase indicates that the ground state is adiabatically connected with the “irreducible cluster state”. 
We numerically calculate the Berry phase of the numerically obtained state and compare with that of the generalized VBS. 

\subsection{Generalized VBS state}
In the EBHM with ${\bar n}=1$ and three-body constraint, the VBS of the HI phase was proposed by Berg, {\it et.al.} \cite{Berg2008} as 
\begin{eqnarray}
|\Psi_{HI}\rangle = C\prod^{L-1}_{j=0}(b^{\dagger}_{j}+b^{\dagger}_{j+1})|0\rangle,
\label{HI_wave_function}
\end{eqnarray}
where $C$ is a normalization constant. 
This state $|\Psi_{HI}\rangle$ are made up of the boson of the ``bonding state", $(b^{\dagger}_j+b^{\dagger}_{j+1})/\sqrt{2}$ 
and can be regarded as the irreducible cluster state under the bond centered inversion symmetry for ${\bar n}=1$ case.
The state $|\Psi_{HI}\rangle$ captures the typical properties of the SPT phase \cite{Berg2008,Stumper,Yang} such as the entanglement spectrum, which has been extensively studied and confirmed \cite{Ejima2014}. The bonding state corresponds to spin $1/2$ singlet in the VBS state \cite{Hatsugai2006}.

Here, as for the EBHM with dimerization, we propose a generalized VBS 
\begin{eqnarray}
|\Psi^{p,q}\rangle = C\prod^{(L-1)/2}_{m=0}(b^{\dagger}_{2m}+b^{\dagger}_{2m+1})^{p}(b^{\dagger}_{2m+1}+b^{\dagger}_{2m+2})^{q}|0\rangle,\nonumber\\
\label{SPT_wave_function}
\end{eqnarray}
where $p+q=2{\bar n}$. 
The bonding state resides on each link, the numbers of which are $p$ and $q$ for $J_1$- and $J_2$-links.
This state $|\Psi^{p,q}\rangle$ is short-range entangled [The entanglement properties are investigated in Appendix B].   
The states $|\Psi^{p,q}\rangle$ can be a typical SPT state appeared in the EBHM of Eq.~(\ref{EBHM}) 
if the local boson Hilbert space is truncated up to $|2{\bar n}\rangle$. 
This $|\Psi^{p,q}\rangle$ is also analogue of the VBS states in the generic spin-$S$ chains, which has been extensively studied before \cite{Katsura2007,Hirano2008,Takayoshi2015,Miyakoshi2016,Moudgalya2018_1,Moudgalya2018_2}. 
The state $|\Psi^{p,q}\rangle$ has the characteristic properties (the presence of the edge states by cutting the system and the quantized Berry phase as shown later.)
In the next subsection, we shall numerically demonstrate the ground states share the properties with $|\Psi^{p,q}\rangle$.

\subsection{Berry phase characterization}

As for the higher-integer spin-$S$ systems and its fermionic analogue \cite{Hatsugai2006,Hirano2008,Mila,Fubasami}, the Berry phase has been employed for the detection of the various SPT states as a local topological order parameter. Especially the exact analytical calculation of the Berry phase for the generic VBS states has been given \cite{Katsura2007}. 

The $Z_2$ Berry phase is given by introducing a local twist \cite{Hatsugai2006}, 
$J_2(e^{i \theta }b^{\dagger}_0 b_{L-1}+e^{-i \theta }b^{\dagger}_0 b_{L-1})$, ($e^{i \theta }\in S^1$, $\theta\in (\pi,\pi]$), as 
\begin{eqnarray}
i\gamma=\int_{S^1} A_{\theta}(\theta)d\theta ,
\label{BP}
\end{eqnarray}
where $A_{\theta}(\theta) =
\langle G(\theta )|\partial_{\theta}G(\theta)\rangle$ 
and $|G(\theta)\rangle$ is the (unique gapped) ground state of $H_{EBH}(\theta)$. 

If the system is bond centered symmetric and its ground state is gapped unique, 
the Berry phase $\gamma$ is quantized by $0$ or $\pi$. 
The value of $\gamma$ does not change as long as the gap is open. 
The Berry phase for the VBS state $|\Psi^{p,q}\rangle$ is $\gamma=q\pi$ (mod $2\pi$), corresponding to the number of the bonding state $(b^{\dagger}_{L-1}+b^{\dagger}_{0})/\sqrt{2}$ under mod $2\pi$ [See Appendix B].

Then, if a gapped ground state is deformed into the VBS state $|\Psi^{p,q}\rangle$ without gap closing, the ground state has the same value of the Berry phase. 
In the next subsection, we numerically demonstrate that some unique gapped ground states have the same Berry phase as that of the VBS state $|\Psi^{p,q}\rangle$.

\subsection{Numerical demonstration of the various SPT phases}
In what follows, let us focus on bulk properties. 
For the EBHM, we numerically calculate the Berry phase, ES and entanglement entropy (EE) 
and compare them to the properties of the VBS state $|\Psi^{p,q}\rangle$. 

\begin{figure}[t]
\begin{center} 
\includegraphics[width=9cm]{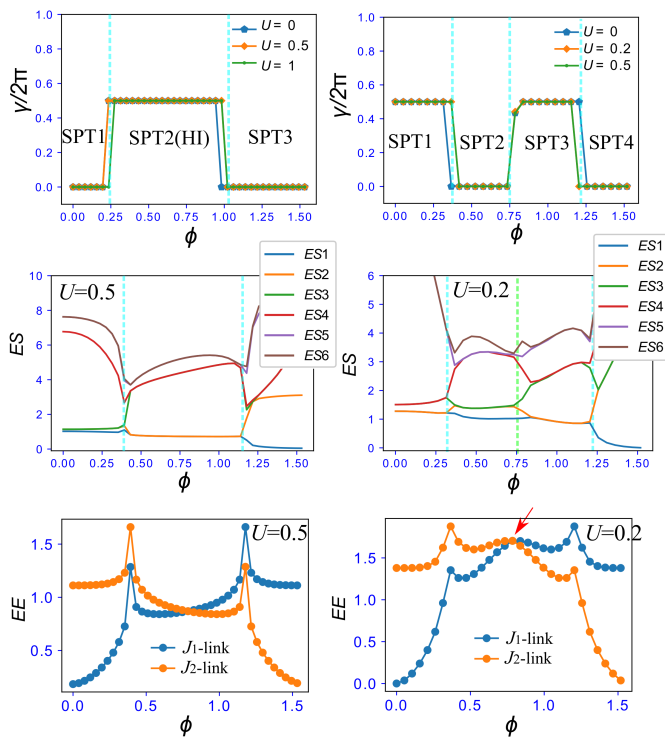} 
\end{center} 
\caption{Berry phase $\gamma$: (a) $\bar{n}=1$, three-body constraint, 
(b) $\bar{n}=2/3$, four-body constraint. 
For (a) and (b) data, the system sizes are $L=10$ and $L=8$.
(c) Six lowest ESs for $J_2$-link, labeled by $EE1 \sim EE6$. The system is set in $U=0.5$ and $\bar{n}=1$.
(d) Six lowest ESs for $J_2$-link, labeled by $EE1 \sim EE6$. The system is set in $U=0.2$ and $\bar{n}=3/2$.
(e) EE for $J_2$-link. The system is set in $U=0.5$ and $\bar{n}=1$.
(f) EE for $J_2$-link. The system is set in $U=0.2$ and $\bar{n}=3/2$.
In ES and EE calculations by iDMRG, the system size is $L=32$.}
\label{Fig1}
\end{figure}

Let us first discuss the case, ${\bar n}=1$ and adapt three-body constraint, where the local boson Hilbert space is limited to $|0\rangle$, $|1\rangle$ and $|2\rangle$.
We also set the parameters of the EBHM as 
$J_1=J\sin\phi$, 
$J_2=J\cos\phi$, 
$V_{j}=V$, and the NN interaction is uniform. 
We set $J=V=1$ \cite{SF_come} and restrict $U=0$, $0.5$ and $1$ for ${\bar n}=1$ \cite{U_dependence}. 
Varying the dimerization parameter $\phi$ between $0$ and $\pi/2$, we calculate the Berry phase $\gamma$ by the exact diagonalization \cite{Quspin}, ES and EE by the infinite-system density matrix renormalization group (iDMRG) algorithm by the TeNPy package \cite{TeNPy}. 

Figure.~\ref{Fig1} is the Berry phase $\gamma$. 
The results indicate the topological phase transitions by varying the dimerization $\phi$.  
In Fig.~\ref{Fig1} (a) the gap remains open except for the transition points. 
We find the three SPT phases labeled as SPT1, SPT2 (HI), and SPT3. 
These phases have same Berry phase as that of $|\Psi^{0,2}\rangle$, $|\Psi^{1,1}\rangle$, and $|\Psi^{2,0}\rangle$.

We further calculate the ES by cutting the $J_2$-link. 
The structure of the ES identifies whether the gapped ground state is trivial or nontrivial that corresponds to the appearance of edge states. 
In Fig~\ref{Fig1} (b), we plot the six lowest ESs as changing $\phi$ with $U=0.5$. There are three regimes: 
For $0\lesssim \phi \lesssim \pi/8$, the lowest three ESs are nearly three-fold degenerate, 
for $\pi/8 \lesssim \phi \lesssim 3\pi/8$, the two lowest ES are degenerate, which is consistent to the previous results \cite{Ejima2014}, and  
for $3\pi/8 \lesssim \phi \lesssim \pi/2$, the lowest ES is isolated.
These low-lying structures are qualitatively consistent with the structure of the ES obtained by the VBS states, $|\Psi^{2,0}\rangle$ and $|\Psi^{1,1}\rangle$ [See Appendix A]. Also, from the obtained ES the entanglement entropy (EE) by cutting the $J_1$- or $J_2$-links are shown in Fig.~\ref{Fig1} (c). 
The two peaks of the EE indicate the topological phase transitions.
The numerical results of $\gamma$ and the ES indicate the obtained states in the EBHM have the same properties as that of the VBS Eq.~(\ref{SPT_wave_function}). 

We next consider the denser case ${\bar n}=3/2$ with the four-body constraint, where the local boson Hilbert space is truncated up to $|3\rangle$. Here, we set a modulated NN interaction: $V_{j\in odd}=J_1,\:\:V_{j\in even}=J_2$, 
the other parameter forms are the same as ${\bar n}=1$ case. 
Here, we restrict the case with $U=0$, $0.2$ and $0.5$ \cite{U_dependence2}.

The Berry phase $\gamma$ is shown in Fig.~\ref{Fig1} (d). 
There are four SPT phases, where we label them as SPT1, SPT2, SPT3, and SPT4. 
These phases have same Berry phases as that of $|\Psi^{0,3}\rangle$, $|\Psi^{1,2}\rangle$, $|\Psi^{1,2}\rangle$ and $|\Psi^{0,3}\rangle$. 
It is noted that the transition between SPT2 and SPT3 occurs at $\phi=\pi/4$, the translational point ($J_1=J_2$), this is expected by the Lieb-Schultz-Mattis type argument for half-integer spin-$S$ systems with translational symmetry \cite{Hirano2008}. 
It indicates that at $\phi=\pi/4$, the ground state of ${\bar n}=3/2$ is gapless while that of ${\bar n}=1$ gapped.
Figure~\ref{Fig1} (e) is the six lowest ESs as changing $\phi$ with $U=0.2$. 
There are four regimes: 
For $0\lesssim \phi \lesssim \pi/8$, the two pairs of the two-fold degenerate ESs appear, 
for $\pi/8 \lesssim \phi \lesssim \pi/4$, single lowest ES and the two-fold degenerate next lowest ES appear,   
for $\pi/4\lesssim \phi \lesssim 3\pi/8$, the lowest two ES are two-fold degenerate, and for $3\pi/8 \lesssim \phi \lesssim \pi/2$, the lowest ES is isolated. These low-lying structures are similar to the structure of the ES expected by the VBS states, $|\Psi^{3,0}\rangle$ and $|\Psi^{2,1}\rangle$ [See Appendix A]. 
Also, as shown in Fig.~\ref{Fig1} (f), the behavior of EEs obtained by cutting the $J_1$- or $J_2$-links supports the presence of the phase transitions. The EEs exhibit peaks at the transition point. 
In particular, though capturing the phase transition around $\phi=\frac{\pi}{4}$ by the ES is subtle, the EE shows a weak bend [the red arrow in Fig~\ref{Fig1} (f)] where the differentiation seems to be discontinuous. 
This behavior implies the topological phase transition.   

Summarizing the behavior of $\gamma$ and ESs, each numerically obtained states for the cases ${\bar n}=1$ and $3/2$, are consistent to the VBS $|\Psi^{p,q}\rangle$.
It suggests the ground states (numerically obtained) in the EBHM can be adiabatically connected to the VBS $|\Psi^{p,q}\rangle$.

In addition, we calculated the ${\bar n}=3/2$ case with uniform $V_j=V$, which is a suitable setting in real experiments, as shown in Appendix C. 
There, the Berry phase $\gamma$ also indicates similar phase structure to Fig.~\ref{Fig1} (d), 
which implies that there are various SPT phases for the uniform $V$.
As for the structure of the low-lying ES,
the deviation from the structure of the ES in the VBS states $|\Psi^{p,q}\rangle$ is larger than that of the modulated $V_j$ case. 
However, the clustering feature of the low-lying ESs remains.

\section{Topological charge pump} 
Based on the presence of the various SPT phases, 
one can extend parameter space by introducing symmetry breaking terms protecting the SPTs, whereas the global $U(1)$-symmetry remains respected. 
Then, one can set a path connecting different SPT phases without closing the gap \cite{Berg2011,KH2020,KH2021}. 

As the simplest symmetry breaking term, we introduce a staggered potential 
\begin{eqnarray}
V_d=\sum^{L-1}_{j=0}(-1)^{j}\Delta(t)n_{j}, 
\label{Vp}
\end{eqnarray} 
where $\Delta (t)$ is set in later. $V_d$ breaks the bond centered inversion symmetry. 
For the Hamiltonian $H_{EBH}+V_d$, one sets the time dependent pump path, $\phi\to \phi(t)=\phi_{i}+\phi_m [1-\cos(2\pi t/T)]/2$ and $\Delta(t)=\Delta_0 \sin\phi(t)$, where $t$, $T$ and $\Delta_0$ are time, the period of the pump, and the strength of the staggered potential, respectively. 
For a suitable choice of the parameters, the pump path wraps the gapless transition points of the SPT phase, where the ground state of $H_{EBH}+V_d$ with periodic boundary on the pump path remains to be unique and gapped \cite{Berg2011,KH2020,KH2021}. 
For $t=0$ and $T/2$, the symmetry is recovered where the SPT phases are defined.
In what follows, $\Delta_0=-1$. 
\begin{figure}[t]
\begin{center} 
\includegraphics[width=8.5cm]{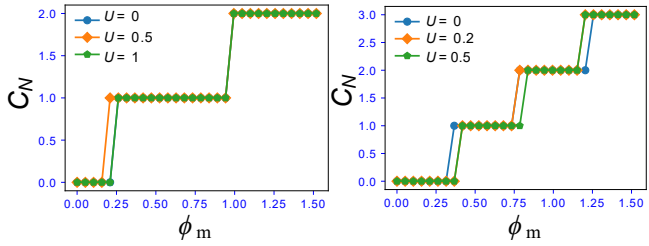} 
\end{center} 
\caption{(a) Chern number $C_N$ of the bulk pumping as increasing $\phi_m$ with $\phi_i=0$. 
We imposed $\bar{n}=1$, three-body constraint.
(b) $\bar{n}=3/2$ case with $\phi_i=0$. 
The four-body constraint is imposed. 
The system size is $L=10$ for (a) and $L=8$ for (b).}
\label{Fig2}
\end{figure}

\subsection{Bulk pump}
We numerically demonstrate the presence of topological charge pumps in the bulk. 
The bulk topological pump is characterized by the many-body Chern number \cite{Niu1985}
\begin{eqnarray}
C_N=\frac{i}{2\pi T}\int_{T^2}B(\theta,t)d\theta dt,
\label{CN}
\end{eqnarray}
where $B(\theta,t)=\partial_{\theta}\langle \Psi(\theta,t)|\partial_t|\Psi(\theta,t)\rangle-\partial_{t}\langle \Psi(\theta,t)|\partial_{\theta}|\Psi(\theta,t)\rangle$, $|\Psi(\theta,t)\rangle$ is a unique ground state of the Hamiltonian $H_{EBH}(\theta,t)+V_d(t)$ and $T^2=[-\pi,\pi)\times [0,T)$. 
$C_N$ corresponds to the total pumped charge per one pump cycle \cite{Hatsugai2016,KH2020,KH2021,KKH2021}. 

For ${\bar n}=1$ case with three-body constraint, 
we calculated $C_N$ \cite{FHS2005} as varying $\phi_m$ for $\phi_i=0$ and $U=0$, $0.5$ and $1$ as shown in Fig.~\ref{Fig2} (a). 
$C_N$ is quantized and quantum phase transitions characterized by $C_N$ appear, which are the plateau transitions of the topological pump. 
The transition of the pump occurs when the path is passing through the transition points of the SPT characterized by the Berry phases $\gamma$. 
This implies that the transition point of the SPT is a topological obstruction in the $\Delta(t)$-$\phi$ parameter space, which induces the quantization of $C_N$. 

As for the case ${\bar n}=3/2$ with four-body constraint, the plots of $C_N$ for $\phi_i=0$ and $U=0$, $0.2$ and $0.5$ are shown in Fig.~\ref{Fig2} (b). 
Similar quantum plateau transitions appear with several plateaus of $C_N$ reflecting the existence of the SPT phases. 
Summarizing the results of ${\bar n}=1$ and $3/2$ cases, the EBHM exhibits various topological charge pumps in the bulk. 

In addition, for ${\bar n}=3/2$ case, a uniform $V$ case is also treated in Appendix C. The same plateau transitions appear.

\begin{figure}[t]
\begin{center} 
\includegraphics[width=8.5cm]{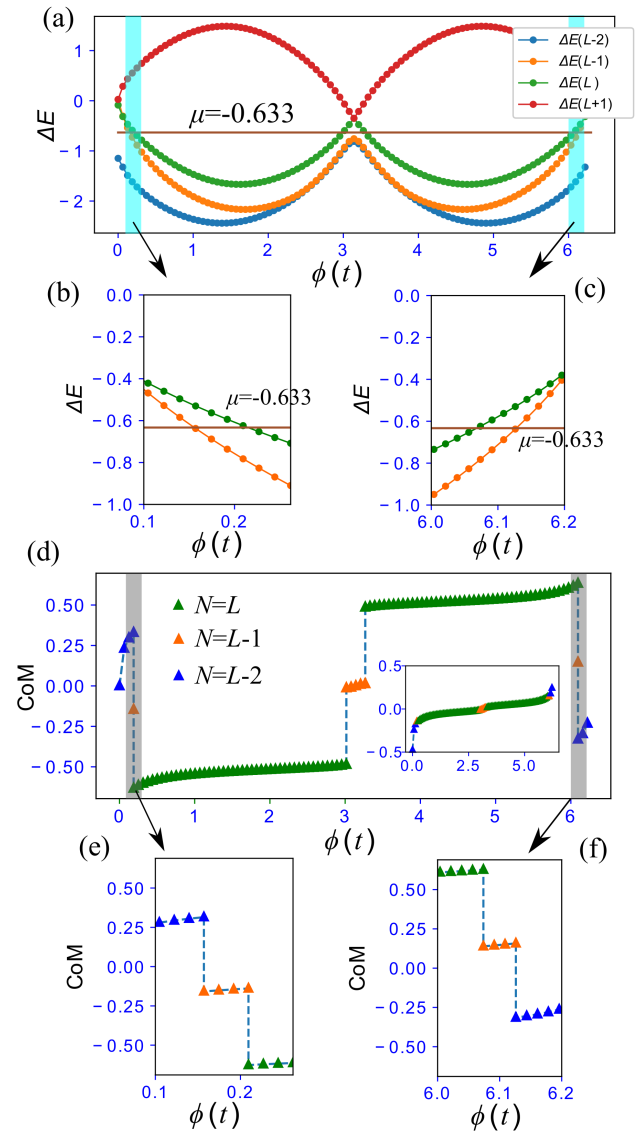} 
\end{center} 
\caption{(a) Excitation energies $\Delta E(N)$ during pumping.  
The closeup data of $\Delta E(N)$ around $t= 0.15$ and $6.1$ are shown in (b) and (c). 
(d) Whole behavior of the CoM with $\mu=-0.633$. 
Six jumps occur at $t\sim 0.15$ $\pi$ and $6.1$. The inset shows the connected data of the CoM obtained by eliminating the jump.
The closeup data of $\Delta (N)$ around $t= 0.15$ and $6.1$ are shown in (e) and (f). 
Each jumps takes $|\Delta P|\sim 0.5$ except for the sigh.  
The system size is $L=64$. We set $U=0.5$.
}
\label{Fig3}
\end{figure}

\subsection{Open boundary case and bulk-edge correspondence}
Next let us discuss the systems with open boundary condition. 
To characterize the topological pump with open boundary, 
we calculate the CoM given by 
\begin{eqnarray}
P(t)=\frac{1}{L}\sum^{L-1}_{j=0}(j-j_0)\langle \Psi(t)|n_j|\Psi(t)\rangle,
\label{CoM}
\end{eqnarray}
where $j_0=(L-1)/2$ and $|\Psi(t)\rangle$ is a ground state of the system at time $t$. 
The gragh of $P(t)$ during pumping is 
given by pairwise continuous parts with discontinuities (jumps). 
The jump of $P(t)$ is defined by $\Delta P(t_i)=P(t_i-0)-P(t_{i}+0)$, where $t_i$ is a time at the discontinuities. 
In the smooth part, we can define the time derivative of $P(t)$, $\partial_t P(t)$, which corresponds to the bulk current at $t$. 
Sum of the integral over the continuous part, $\sum_{i}\int^{t_{i+1}}_{t_{i}}\partial_t P(t) dt$ gives the total pumped charge denoted by $Q_b$, which corresponds to $C_N$, $Q_{b}=C_N$ \cite{Hatsugai2016}.
On the other hand, $\Delta P(t_i)$ is induced by the creation or annihilation of the left or right edge states. 
Due to the periodicity of $P(t)$, $P(t)=P(t+T)$, 
there is a relation between the total sum of jump $\Delta P(t_i)$ and $C_N$ \cite{Hatsugai2016}
\begin{eqnarray}
C_N+\sum_{i}\Delta P(t_i)=0.
\label{BEC}
\end{eqnarray} 
This is the bulk-edge correspondence of the topological pump.  
In the following, we shall calculate the behavior of the CoM in detail and verify the validity of Eq.~(\ref{BEC}) in the EBHM. 
To this end, we employed the finite density matrix renormalization algorithm by using TeNPy \cite{TeNPy}. 

\begin{figure}[t]
\begin{center} 
\includegraphics[width=8.5cm]{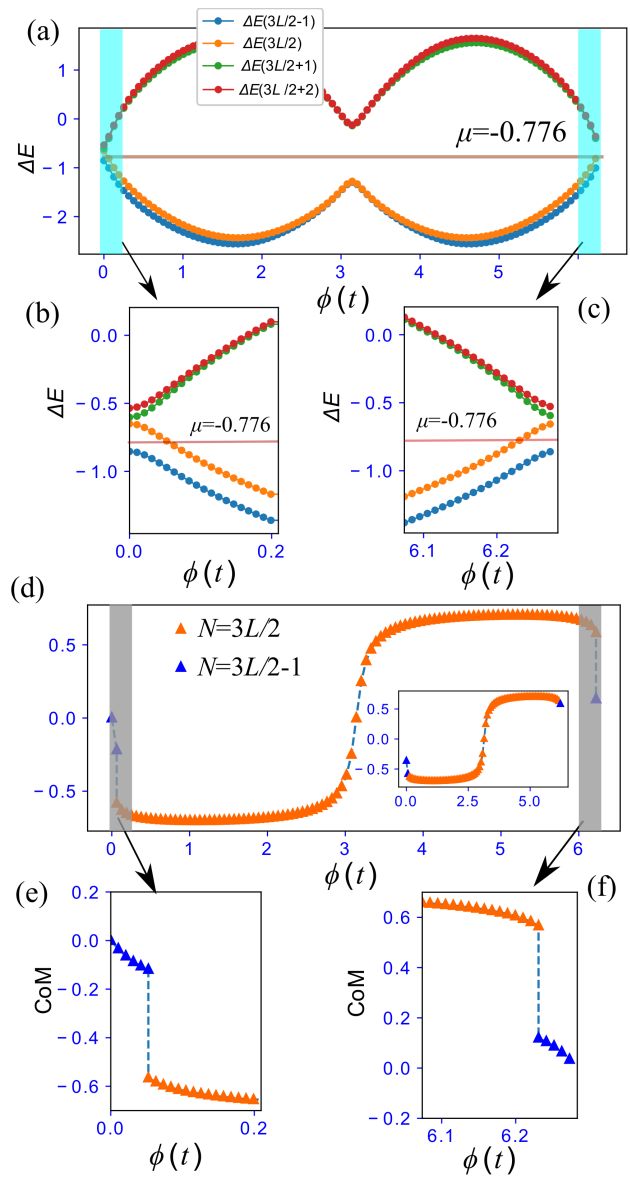} 
\end{center} 
\caption{
(a) Excitation energies $\Delta E(N)$ during the pumping.  
The closeup data of $\Delta E (N)$'s around $t= 0.1$ and $6.2$ are shown in (b) and (c). 
(d) Whole behavior of the CoM with $\mu=-0.776$. 
Two jumps occur at $t\sim 0.1$ and $6.2$. 
The inset shows the connected data of the CoM obtained by eliminating the jump.
The closeup data of $\Delta E (N)$'s around $t=0.1$ and $6.2$ are shown in (e) and (f). 
Each jumps of the CoM takes $\Delta P \sim -0.5$.  
The system size is $L=48$. We set $U=0$.
}
\label{Fig4}
\end{figure}
In the following, we focus on the grand canonical ensemble, that is, we assume that the system touches particle reservoir. 
We add the chemical potential term denoted by $H_{cp}=-\mu\sum^{L-1}_{j=0}n_j$ to the system as $H_{EBH}+V_d+H_{cp}$. 
During pumping, the energy and total particle number of the system vary. 
The ground state is a minimum energy state of the $H_{EBH}+V_d+H_{cp}$. 
To determine the total particle number during the pumping, 
we plot the spectral flow of the excitation energy, $\Delta E(N)=E_{0}(N)-E_{0}(N-1)$, where $E_{0}(N)$ is the ground state energy of $H_{EBH}+V_d$ with $N$ particles. 
We calculate the total energies of the system with $N$ and $N-1$ particles. 
These are given by $E_{0}(N)-\mu N$ and $E_{0}(N-1)-\mu(N-1)$. 
Here, the number of particles can be determined by which energy is lower: 
if the particle number $N_c$ satisfies $\Delta E(N_c)<\mu< \Delta E(N_c-1)$, 
the $N_c$ particle system becomes the ground state. 

As the first typical case, we focus on the case, ${\bar n}=1$ with $\phi_i=\pi/24$ and $\phi_m=5\pi/24$, 
where the pump path starts from the SPT1 phase and then passing through the SPT2(HI) phase at $t=T/2$. 
This path has $C_N=1$ in the bulk. 
For this pump, several $\Delta E(N)$ around $N={\bar n}L=L$ are plotted in Fig.~\ref{Fig3} (a). 
At $t=0$ and $T/2$, 
some $\Delta E$'s almost cross but in the finite-size system, the overlap of exponentially localized edge states induces small energy splitting. This small gap is generated by the effective interaction between the left and right edge states, that scales $\exp{[-L/\xi]}$ ($\xi$ is the correlation length). 
Hence, for $L\to \infty$, the gap closes \cite{Kennedy,edge_ene}.
In the periodic system, such low energy states do not exist. 
Noting that, generally, 
symmetries fix the energy of edge states as shown in \cite{Hatsugai2006, KH2020}, 
where particle-hole symmetry fixes edge states to zero-energy.
However, due to the absence of the particle-hole symmetry in the EBHM, 
the energies of edge states are not strictly fixed in the HI phase \cite{Berg2008,Stumper}. 

We here set $\mu=-0.633$. 
The energy lines $\Delta E(L)$ and $\Delta E(L-1)$ cross $\mu$-line at $t\sim 0.15$ and $6.1$, the closeup data are shown in Fig.~\ref{Fig3} (b) and (c), where the number of particle of the ground state changes.
Selecting the suitable particle number determined by the data of $\Delta E(N)$'s, we plot the spectral flow of the CoM in Fig.~\ref{Fig3} (d)-(f). 
During the pump, the particle number changes between $N=L-2$ and $L$ where the CoM jumps. 
The whole plot of the CoM is shown in Fig.~\ref{Fig3} (d), where the CoM jumps at $t\sim 0.15, \pi$ and $6.1$, where the particle number changes by one. 

These jumps are induced by the creation or annihilation of the left or right edge states. 
We further calculate the CoM around $t= 0.15$ and $6.1$. The closeup data of the CoM are shown in Fig.~\ref{Fig3} (e) and (f). 
We find that clear two jumps appear and each jumps take approximately $\Delta P\sim \pm 0.5$ (The sigh depends on the creation or annihilation of left or right edge states \cite{Hatsugai2016}). 
This indicates that there are two edge states on both edges and their edge states induce the jump of CoM by the creation or annihilation.
In the presence of these multiple edge states, 
it is crucial to consider such a grand canonical ensemble to include the contribution of edge states to the CoM. 
These observations of the jumps of the CoM are consistent with the discussion by \cite{Hatsugai2016}. 
Such a contribution of the edge states is not captured 
by the calculation of the CoM with fixed particle number (See Appendix D, where around $t\sim 0$ the two edge states exchange from right to left almost at the same time. However, it is difficult to observe each contribution of two edge states.)

In the whole behavior of the CoM in Fig.\ref{Fig3} (d), if we connect the smooth parts as shown in the inset in Fig.\ref{Fig3} (d), then we see $Q_b\sim 1$.  Also, the total sum of $\Delta P$ for one pump cycle. We conclude $-\sum_{t_i}\Delta P(t_i)\sim 1$. It quantizes as $-\sum_{t_i}\Delta P(t_i)=1$ for $L\to \infty$. 
In conclusion, we confirmed the bulk-edge-correspondence of the topological pump of Eq.~(\ref{BEC}) in the EBHM.

As the second typical case, we consider the case, ${\bar n}=3/2$ with $\phi_i=\pi/3$ and $\phi_m=\pi/8$, where starting with the SPT3 phase, the path crosses the SPT4 phase at $t=T/2$, where the pump has $C_N=1$ in the bulk since the path goes around the single gapless phase transition point. 
The same calculations as the case, ${\bar n}=1$ are carried out. For this pump, the several $\Delta E(N)$ around $N=L$ are plotted in Fig.~\ref{Fig4} (a). 
We set $\mu=-0.776$. 
Here, the energy line $\Delta E(L)$ crosses the $\mu$-line at $t\sim 0.1$ and $6.2$, the closeup data are shown in Fig.~\ref{Fig4} (b) and (c). 
Here, note that $\Delta E(3L/2+1)$ and $\Delta E(3L/2)$ almost cross at $t=0$ but a small finite gap appears since the overlap of exponentially localized edge states induces small energy splitting.

Selecting the suitable particle number determined by the data of $\Delta E(N)$, we plot the spectral flow of the CoM in Fig.~\ref{Fig4} (d)-(f). The whole plot of the CoM is shown in Fig.~\ref{Fig4} (d), where the CoM jumps twice at $t\sim 0.2$ and $6.2$. 
The inset in Fig.~\ref{Fig4} (d) shows the connected data of the CoM. This behavior indicates $Q_b\sim 1$. The closeup data of the CoM around $t=0.1$ and $6.2$ are shown in Fig.~\ref{Fig4} (e) and (f), each jumps of the CoM take approximately $\Delta P\sim -0.5$ to give $-\sum_{i}\Delta P(t_i)\sim 1$. It quantizes as $-\sum_{t_i}\Delta P(t_i)=1$ for $L\to \infty$. Therefore, we also confirmed the bulk-edge-correspondence of Eq.~(\ref{BEC}). The behavior of the CoM for ${\bar n}=3/2$ with fixed particle number is also shown in Appendix D, where around $t=0$ the edge states exchange from right to left almost at the same time, but each contributions of the edge state to the jump cannot be captured.


\section{Conclusion}
In this work, we demonstrated that the interplay of the hopping dimerization, NN repulsive interactions, mean density, and the truncation of the local boson Hilbert space induces various SPT phases and topological charge pumps. 
 
The various SPT phases in the EBHM are analogous to the various VBS states 
emerged in the generic spin-$S$ dimerized spin model. 
We proposed the generalized VBS states in the bosonic system to capture the properties of the SPT of the EBHM.  
To characterize the SPT phases in the bulk, we employed the $Z_2$ Berry phase. 
The calculation gives phase boundaries even for a small finite-size system. 
Also, we investigated the low-lying structure of ESs and observed that the degenerate structure also characterizes the bulk SPT phases. 
{In our numerical calculations, the structure of ESs is consistent with that expected by the VBS states $|\Psi^{p,q}\rangle$. These numerically obtained ground states in the EBHM can be adiabatically connected to the VBS states.} 

Furthermore, based on the presence of various bulk SPT phases, 
we have realized various topological charge pumps in the EBHM. 
The plateau transitions of the bulk topological pump appear. 
We next focus on the topological pump in the system with open boundary condition by employing DMRG. 
In particular, the CoM is investigated in detail for the grand canonical ensemble. 
In this situation, the contributions of each edge state to the jump of the CoM are separately captured. 
We numerically confirm that these bosonic pumps obey the bulk-edge-correspondence.
 
We finally comment that the EBHM in this work is feasible in a real experimental system such as cold atoms with dipole-dipole interactions \cite{Baier} and also drawing the detailed global phase diagram and quantifying the detailed dependence of the truncation of the local Hilbert space will be interesting topics as future work.

\section{ACKNOWLEDGMENTS}
The authors thank K. Kudo for valuable discussions. The work is supported by JSPS
KAKEN-HI Grant Number JP17H06138 and JP21K13849 (Y.K.).

\renewcommand{\thesection}{A\arabic{section}} 
\renewcommand{\theequation}{A\arabic{equation}}
\setcounter{equation}{0}
\widetext
\section*{Appendix A: Entanglement spectrum for the generalized VBS states}
Motivated by the dimerized spin-$S$ chain \cite{Hirano2008} and a previous works \cite{Berg2008,Yang}, we expect that the essential properties of SPT phases of the EBHM can be captured by the generalized states 
\begin{eqnarray}
|\Psi^{p,q}\rangle = C\prod^{L/2-1}_{m=0}(b^{\dagger}_{2m}+b^{\dagger}_{2m+1})^{p}(b^{\dagger}_{2m+1}+b^{\dagger}_{2m+2})^{q}|0\rangle,
\label{SPT_wave_function}
\end{eqnarray}
where $p+q=2{\bar n}$ (${\bar n}$ is mean particle density) and $C$ is a normalized constant. 
Here we can write down the MPS form for some simple ($p$,$q$) cases. 
From the MPS form we can extract the structure of the low-lying ESs and the EE for the state $|\Psi^{p,q}\rangle$. 
These properties of the ESs and EE can be compared to those of the ESs and EEs in the SPT states obtained by the numerical simulations.

\subsection{$\bar{n}=1$ and three-body constraint case}
First example is the dimerized case $J_1=1$ and $J_2=0$. 
The SPT phase can be captured by 
\begin{eqnarray}
|\Psi^{2,0} \rangle = C_{20}\prod^{L/2-1}_{m=0}(b^{\dagger}_{2m}+b^{\dagger}_{2m+1})^{2}|0\rangle,
\label{SPT_N=1_dimer}
\end{eqnarray}
where $C_{20}$ is a normalization constant.

Since the system is in the dimerized limit, we focus on the two site, where the dimers reside. 
For the two site system, by employing the singular value decomposition, the state is transformed into the two site MPS form 
\begin{eqnarray}
\frac{1}{2\sqrt{2}}(b^{\dagger}_{j_1}+b^{\dagger}_{j_2})^{2}|0\rangle=\sum_{j_1,j_2=0,1,2}\mathrm{Tr}\biggl[\Gamma^{A}_{j_1}\Lambda\Gamma^{B}_{j_2}\biggr]|j_1j_2\rangle,
\label{two_site_dimer_MPS}
\end{eqnarray}
where 
\begin{eqnarray}
\Gamma^{A}_{0}&=&[0,0,-1],\:\: \Gamma^{A}_{1}=[0,-1,0],\:\: \Gamma^{A}_{2}=[1,0,0],\\
\Lambda_B&=&\left[
\begin{array}{ccc}
1/2&0&0\\
0&1/\sqrt{2}&0\\
0&0&1/2
\end{array}
\right],\:\:
\Gamma^B_{0}=\left[
\begin{array}{c}
0\\
-1\\
0
\end{array}
\right],\:\:
\Gamma^B_{1}=\left[
\begin{array}{c}
0\\
0\\
-1
\end{array}
\right],\:\:
\Gamma^B_{2}=\left[
\begin{array}{c}
1\\
0\\
0
\end{array}
\right].
\end{eqnarray}
The matrix $\Lambda$ determines the ES and EE when the double dimers are cut. 
The three ESs of $\Lambda$ are nearly three-fold and the EE is given by $\frac{3}{2}\ln 2$. 
From the two site MPS form, the VBS state with $(p,q)=(2,0)$ is also written by
\begin{eqnarray}
|\Psi^{2,0}\rangle &=&\bigotimes^{L/2-1}_{m=0}\biggl[\sum_{j_1,j_2=0,1,2}\mathrm{Tr}\biggl[\Gamma^{A}_{j_{2m}}\Lambda\Gamma^{B}_{j_{2m+1}}\biggr]|j_{2m}j_{2m+1}\rangle \biggr]. 
\label{SPT_N=1_dimer_v2}
\end{eqnarray}

We next consider the $(p,q)=(1,1)$ case, which is the VBS state for the Haldane insulator (HI),
\begin{eqnarray}
|\Psi^{1,1}\rangle = C_{11}\prod^{L-1}_{m=0}(b^{\dagger}_{m}+b^{\dagger}_{m+1})|0\rangle,
\label{SPT_N=1_HI}
\end{eqnarray}
where $C_{11}$ is a normalization constant. 
For the state $|\Psi^{1,1}\rangle$, we find the following MPS representation 
\begin{eqnarray}
|\Psi^{1,1}\rangle =C^{M}_{11} \sum_{j_{\ell}=0,1,2}\mathrm{Tr}\biggl[\prod^{L-1}_{\ell=0}A^{[j_\ell]}\biggr] |j_{0}\cdots j_{L-1}\rangle,
\label{MPS_N=1_HI}
\end{eqnarray}
where $C^{M}_{11}$ is a normalization constant and 
\begin{eqnarray}
A^{0}=\frac{1}{\sqrt{2}}\left[
\begin{array}{cc}
0&1\\
0&0
\end{array}
\right],\:\:
A^{1}=\frac{1}{\sqrt{2}}\left[
\begin{array}{cc}
1&0\\
0&1
\end{array}
\right],\:\:
A^{2}=\left[
\begin{array}{cc}
0&0\\
1&0
\end{array}
\right].
\end{eqnarray}
It should be noted that the matrices $A^{[\alpha]}$ are different from those of the VBS for the $S=1$ AKLT model.
By considering three site HI with the periodic boundary condition, we can easily confirm that the MPS form surely represents the state $|\Psi^{1,1}\rangle$, 
\begin{eqnarray}
|\Psi^{1,1}\rangle =\sqrt{2}\sum_{j_{0},j_{1},j_{2}=0,1,2}\mathrm{Tr}\biggl[ A^{j_0}A^{j_1}A^{j_3}\biggr] |j_{0}j_1 j_2\rangle \nonumber\\
=\frac{1}{4}(b^{\dagger}_0+b^{\dagger}_1)(b^{\dagger}_1+b^{\dagger}_2)(b^{\dagger}_2+b^{\dagger}_0)|0\rangle.
\label{SPT_N=1_three_site_check}
\end{eqnarray}
Surely, the above MPS with periodic boundary case reproduces the three site periodic system of $|\Psi^{1,1}\rangle$.

For an infinite system size, the MPS of the HI state of Eq.~(\ref{MPS_N=1_HI}) can be transformed into the canonical form \cite{Schollwock2011}. It is possible by imposing a suitable gauge matrix transformation $A^{[j]}\to XA^{[j]}X^{-1}$ \cite{Orus2008,Takayoshi2015}. We can find the unitary matrix $M$. The canonical infinite MPS of the state $|\Psi^{1,1}\rangle$ is written by  
\begin{eqnarray}
|\Psi^{1,1}\rangle =\sum_{j_{\ell}=0,1,2}[\cdots \Lambda \Gamma^{[j_{\ell}]}\cdots] |\cdots j_{\ell}\cdots\rangle,
\label{SPT_N=1_dimer_v2}
\end{eqnarray}
where 
\begin{eqnarray}
\Lambda&=&\left[
\begin{array}{cc}
1/\sqrt{2}&0\\
0&1/\sqrt{2}
\end{array}
\right],\:\:
\Gamma^{[0]}=K\left[
\begin{array}{cc}
0&1\\
0&0
\end{array}
\right],\:\:
\Gamma^{[1]}=K\left[
\begin{array}{cc}
1&0\\
0&1
\end{array}
\right],\:\:
\Gamma^{[2]}=K\left[
\begin{array}{cc}
0&0\\
1&0
\end{array}
\right],\:\:
K=\frac{\sqrt{2}}{\sqrt{1+\sqrt{2}}}.
\end{eqnarray}
This MPS representation is surely the canonical form because the transfer matrix of the MPS satisfies the following canonical condition \cite{Orus2008},
\begin{eqnarray}
\sum_{\alpha,\alpha'}\delta_{\alpha,\alpha'}T^{\alpha,\alpha'}_{\beta,\beta'}=\delta_{\beta,\beta'}, \:\:
\sum_{\beta,\beta'}\delta_{\beta,\beta'}T^{\alpha,\alpha'}_{\beta,\beta'}=\delta_{\alpha,\alpha'}
\end{eqnarray}
where $T^{\alpha,\alpha'}_{\beta,\beta'}$ is the transfer matrix, 
\begin{eqnarray}
T^{\alpha,\alpha'}_{\beta,\beta'}=\sum_{j}[\Lambda\Gamma^{[j]}]^{*}_{\alpha,\beta}[\Lambda\Gamma^{[j]}]_{\alpha',\beta'}
=\sum_{j}[\Gamma^{[j]}\Lambda]^{*}_{\alpha,\beta}[\Gamma^{[j]}\Lambda]_{\alpha',\beta'}.
\end{eqnarray}
The canonical form gives an insight about the qualitative character of the entanglement structure in the system. Especially, the form of $\Lambda$ gives the ES. The ES is two-fold degenerate for each link and the EE is given by $\ln 2$.

\subsection{$\bar{n}=3/2$ and four-body constraint case}
We start with focusing on $(p,q)=(3,0)$ case. 
As for the dimer limit, $J_1=1$ and $J_2=0$ for ${\bar n}=3/2$, we can easily write down the SPT phase, which is the dimerized case $J_1=1$ and $J_2=0$. 
The SPT phase can be captured by 
\begin{eqnarray}
|\Psi^{3,0}\rangle = C\prod^{\infty}_{m=-\infty}(b^{\dagger}_{2m}+b^{\dagger}_{2m+1})^{3}|0\rangle.
\label{SPT_N=1_dimer}
\end{eqnarray}

As the case $(p,q)=(2,0)$, we focus on the two site, where the bonding states reside. 
The two site state is transformed into the two site MPS form \cite{Schollwock2011}
\begin{eqnarray}
\frac{1}{4\sqrt{3}}(b^{\dagger}_{j_1}+b^{\dagger}_{j_2})^{3}|0\rangle=\sum_{j_1,j_2=0,1,2,3}\mathrm{Tr}\biggl[\Gamma^{A}_{j_1}\Lambda\Gamma^{B}_{j_2}\biggr]|j_1j_2\rangle,
\label{two_site_dimer_MPS_3_2}
\end{eqnarray}
where 
\begin{eqnarray}
\:\: \Gamma^{A}_{0}&=&[0,0,-1,0],\:\:\Gamma^{A}_{1}=[0,0,0,-1],\:\: \Gamma^{A}_{2}=[-1,0,0,0],\:\: \Gamma^{A}_{3}=[0,-1,0,0],\\
\Lambda&=&\left[
\begin{array}{cccc}
\sqrt{2}/4&0&0\\
0&\sqrt{6}/4&0&0\\
0&0&\sqrt{6}/4&0\\
0&0&0&\sqrt{2}/4
\end{array}
\right],\:\:
\Gamma^B_{0}=\left[
\begin{array}{c}
0\\
0\\
0\\
-1
\end{array}
\right],\:\:
\Gamma^B_{1}=\left[
\begin{array}{c}
0\\
0\\
-1\\
0
\end{array}
\right],\:\:
\Gamma^B_{2}=\left[
\begin{array}{c}
0\\
-1\\
0\\
0
\end{array}
\right],\:\:
\Gamma^B_{3}=\left[
\begin{array}{c}
-1\\
0\\
0\\
0
\end{array}
\right].
\end{eqnarray}
From the matrix $\Lambda$, when the triple bonding states are cut, 
the low-lying ESs exhibit two set of two-fold degenerate, and the EE is given by $3\ln 2-\frac{3}{4}\ln 3$.

The structure of the low-lying ES for the $(p,q)=(2,1)$ state can be expected from the structure of the low-lying ESs of $(p,q)=(2,0)$ and $(1,1)$ cases. The state $|\Psi^{2,1}\rangle$ has double bonding states and single one at $J_1$- and $J_2$-links. 
Hence, if we cut the even link of the state $|\Psi^{2,1}\rangle$, the ES is the same structure of the $(p,q)=(2,0)$ case, the three low-lying ESs are nearly three-fold. On the other hand, if we cut the $J_2$-link of the state $|\Psi^{2,1}\rangle$, the ES is the same structure of the $(p,q)=(1,1)$ case, the two low-lying ESs are degenerate. Hence, we expect that the EEs are given by $\frac{3}{2}\ln 2$ and $\ln 2$ for the $J_1$- and $J_2$-links.

\section*{Appendix B: Berry phase characterization for SPT phase}
We shall show that the generalized VBS states has the quantized Berry phase if one introduces a twist, which can be introduced by attaching the phase to a hopping term, in principle. 

We start with the HI state. 
The HI state described by Eq.~(\ref{SPT_N=1_HI}) under the twist $\theta$ for the link between $j=L-1$ and $j_{0}$ can be expressed by
\begin{eqnarray}
|\Psi^{1,1}(\theta)\rangle = C_{11}(b^{\dagger}_{L-1}+e^{i\theta}b^{\dagger}_{0})\prod^{L-2}_{m=0}(b^{\dagger}_{m}+b^{\dagger}_{m+1})|0\rangle.
\label{SPT_N=1_HI_twist}
\end{eqnarray}
For the state $|\Psi^{1,1}(\theta)\rangle$, the Berry phase is given by
\begin{eqnarray}
i\gamma =\int^{2\pi}_{0}d\theta \langle \Psi^{1,1}(\theta)|\partial_{\theta}|\Psi^{1,1}(\theta)\rangle.
\label{Berry_phase}
\end{eqnarray}
Here, we take a gauge fixed form, $|\Psi^{1,1}(\theta)\rangle=e^{-i\theta/2}|\Phi_{HI}(\theta)\rangle$ and the Berry phase are given by
\begin{eqnarray}
i\gamma = i\pi +\int^{2\pi}_{0}d\theta \langle \Phi_{HI}(\theta)|\partial_{\theta}|\Phi_{HI}(\theta)\rangle.
\label{Berry_phase_2}
\end{eqnarray}
By following the procedure of Ref.\cite{Katsura2007} 
the state $|\Phi_{HI}(\theta)\rangle$ is written as follows
\begin{eqnarray}
|\Phi_{HI}(\theta)\rangle &=& \cos\frac{\theta}{2}|\Phi_1\rangle+i\sin\frac{\theta}{2}|\Phi_{2}\rangle,\\
|\Phi_1\rangle &=&  C_{11}(b^{\dagger}_{L-1}+b^{\dagger}_{0})\prod^{L-2}_{m=0}(b^{\dagger}_{m}+b^{\dagger}_{m+1})|0\rangle,\\
|\Phi_2\rangle &=&  C_{11}(b^{\dagger}_{L-1}-b^{\dagger}_{0})\prod^{L-2}_{m=0}(b^{\dagger}_{m}+b^{\dagger}_{m+1})|0\rangle.
\label{RHS_Berry_phase}
\end{eqnarray}
Here, the integrant of the second term of the right hand side in Eq.~(\ref{Berry_phase_2}) is 
\begin{eqnarray}
\langle \Phi_{HI}(\theta)|\partial_{\theta}|\Phi_{HI}(\theta)\rangle = 0, 
\label{RHS_inner_product}
\end{eqnarray}
where we used $\langle \Phi_{1}|\Phi_2\rangle=0$. 
Hence, the HI state is characterized by $\gamma=\pi$.

Also, the same calculation can be applied to the general case of $p$ and $q$. The Berry phases for $J_1$ and $J_2$-links are given by $\gamma=p\pi \:(\text{mod}\: 2\pi)$ and $q\pi \:(\text{mod}\:2\pi)$.      

\begin{figure}[t]
\begin{center} 
\includegraphics[width=18cm]{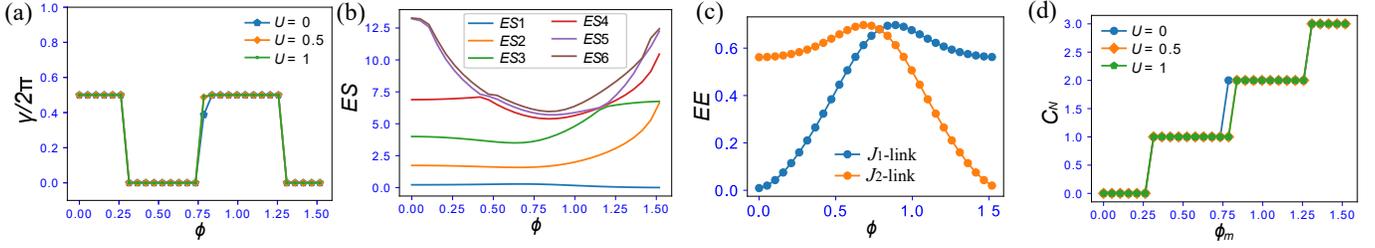} 
\end{center} 
\caption{(a) Berry phase $\gamma$ for $\bar{n}=3/2$ and four-body constraint.
(b) Six lowest ESs for $J_2$-link, labeled by $EE1 \sim EE6$. for $U=0.5$.
(c) EE of $J_1$- and $J_2$-links for $U=0.5$. 
In ES and EE calculations by iDMRG, the system size is $L=48$.
(d) Chern number $C_N$ as increasing $\phi_m$ with $\phi_i=0$.}
\label{FigA1}
\end{figure}

\section*{Appendix C: SPT phases for the uniform $V$ case with ${\bar n}=3/2$ and four-body constraint}
In this appendix, we show the numerical calculation for ${\bar n}=3/2$ with uniform $V$ and four-body constraint. 
The nearly uniform $V$ is feasible for a real experimental system such as dipolar lattice gases trapped in the optical lattice \cite{Baier}. 

The form of the parameter is the same of ${\bar n}=1$ case in Fig.~\ref{Fig1} (a)-(c). 
The Berry phase $\gamma$ is calculated in Fig.~\ref{FigA1} (a) for some values of $U$. 
We found four SPT phases. Figure~\ref{FigA1} (b) is the six-lowest ESs as changing $\phi$ with $U=0.5$. As a whole, the structure of ES gives subtle information about the degeneracy of the ES. 
Moreover, in Fig.~\ref{Fig1} (f), the behavior of EEs for the $J_1$- and $J_2$-links shows no clear peaks. 
Hence, in the uniform $V$ case, 
the Berry phase characterization gives useful information about the bulk phases and their transitions. 

Even for the uniform $V$, we found the topological charge pump 
and the plateau transitions if we set the same pump parameterization as that in the main text. 
The result for $\Delta_0=-1$ and $V=1$ is shown in Fig.~\ref{FigA1} (d).

 
\begin{figure}[h]
\begin{center} 
\includegraphics[width=8cm]{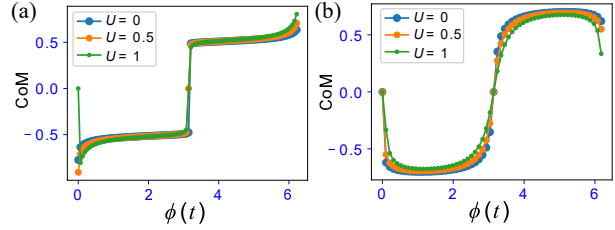} 
\end{center} 
\caption{The CoM behavior with fixed particle number: 
(a) ${\bar n}=1$, $\phi_i=\pi/24$, $\phi_m=5\pi/24$ and the three-body constraint.
(b) ${\bar n}=3/2$, $\phi_i=\pi/3$, $\phi_m=\pi/8$ and the four-body constraint. 
We set $L=64$ and $48$ for (a) and (b).}
\label{FigA2}
\end{figure}

\section*{Appendix D: Behavior of center of mass with fixed particle number}
We show the CoM behavior of the topological pumps with the fixed particle number case. 
We fixed the total particle number with $N={\bar n}L$.
The CoM of the pumps for ${\bar n}=1$ case with $\phi_m=5\pi/24$ are plotted in Fig.~\ref{FigA2} (a). 
We observe a single jump at $t=0$, $T/2$. 
This is different from that in the conventional topological pump in the Rice-Mele model \cite{Greschner2020,KH2020}.
These jumps $\Delta P(t_i)$ at $t=0$, $T/2$ are caused by the creation and annihilation of the left and right edge states at the same time. The sign of $\Delta P(t_i)$ and its amplitude are different.
Even in the fixed particle number case, the total sum of the CoM jump is expected to correspond to $C_N$. Certainly, the numerical result in Fig.~\ref{FigA2}(a) indicates $\sum_{t_i}\Delta P(t_i)\sim 1$ for some values of $U$.
 
Such a CoM behavior is also observed in a topological pump for ${\bar n}=3/2$ case with $\phi_i=\pi/24$ and $\phi_m=5\pi/24$, as shown in Fig.~\ref{FigA2} (b). The CoM jump around $t\sim 0$ is not sharp due to the particle number fluctuations, though roughly we observe $\sum_{t_i}\Delta P(t_i)\sim 1$. Hence, to observe each contribution of the edge states to the CoM and the bulk edge correspondence of the topological pump the grand canonical calculation shown in Fig.3 and 4 is more reasonable.

\endwidetext

\begin{thebibliography}{99}
\bibitem{U_dependence}
It is noted that there is a possibility that larger $U$ induces some topological phase transitions, which leads to more rich phase diagrams on $U-V$ parameter space. This is an interesting future problem.
\bibitem{U_dependence2}
In the numerical calculations of ES, the $U$ dependence is larger than ${\bar n}=1$ case. 
As a tendency, for larger $U$, the structure of the low-lying ESs more deviates from those of the generalized VBS states.
\bibitem{edge_ene}
For ${\bar n}=1$ case as shown in Fig.3 (a), 
such a small finite gap is not observed in our numerical system size $L=64$.
\end{thebibliography}

\end{document}